\documentclass[aps,prb,twocolumn,superscriptaddress,showpacs]{revtex4}
\usepackage{graphicx}
\usepackage{amssymb}
\usepackage{amsmath}

\usepackage{pdfpages}

\begin{document}

\title{Anomalous impurity effect in the heavy-fermion superconductor CeCu$_2$Si$_2$}

\author{Shan Zhao}
\affiliation{School of Physical Science and Engineering, Beijing Jiaotong University, Beijing 100044, China}

\author{Weifu Liu}
\affiliation{School of Physical Science and Engineering, Beijing Jiaotong University, Beijing 100044, China}

\author{Yi-feng Yang}
\thanks{yifeng@iphy.ac.cn}
\affiliation{Beijing National Laboratory for Condensed Matter Physics,  Institute of Physics,
Chinese Academy of Science, Beijing 100190, China}
\affiliation{School of Physical Sciences, University of Chinese Academy of Sciences, Beijing 100190, China}
\affiliation{Songshan Lake Materials Laboratory, Dongguan, Guangdong 523808, China}

\author{Shiping Feng}
\affiliation{Department of Physics, Beijing Normal University, Beijing 100875, China}

\author{Bin Liu}
\thanks{liubin@bjtu.edu.cn}
\affiliation{School of Physical Science and Engineering, Beijing Jiaotong University, Beijing 100044, China}

\date{\today}

\begin{abstract}
Recent observations of a contradictory impurity effect in the heavy-fermion superconductor CeCu$_2$Si$_2$ at ambient pressure have hindered the identification of its pairing symmetry. Here, we perform theoretical analyses with both intraband and interband impurity scatterings for the nodeless $s^\pm$-wave pairing, and report an anomalous nonmonotonic variation of its $T_c$ suppression with the scattering strength. Our results reproduce the prominent reduction of $T_c$ in good agreement with earlier experiments by atomic substitution and explains as well its robustness against electron irradiation. We ascribe the latter to the screening of the interband impurity potential in the strong scattering or unitary region. This resolves the seeming contradiction in different experimental probes and provides an important support to the nodeless $s^\pm$-wave scenario in CeCu$_2$Si$_2$ at ambient pressure. Our theory may be extended to other narrow-band systems such as twisted bilayer graphene and the recently discovered bilayer or trilayer nickelate superconductors, and provide a useful way to distinguish different pairing candidates thereof. 

\end{abstract}

\maketitle
Identifying the pairing symmetry is a central yet extremely challenging issue for unconventional superconductors. One notable example is CeCu$_2$Si$_2$ \cite{Yuan2023}, which was discovered in 1979 as the first unconventional superconductor \cite{Steglich1979} and, for over 30 years, had been widely believed to be of $d$-wave pairing mediated by antiferromagnetic spin fluctuations \cite{Yuan2003,Stockert2004,Eremin2008,Stockert2011,Vieyra2011}. However, in 2014, more elaborate measurements of the specific heat and magnetization on high-quality samples down to 40 mK found surprising thermodynamic evidence for two nodeless gaps \cite{Kittaka2014} and thus questioned the hitherto prevailing $d$-wave scenario. This unexpected observation was subsequently confirmed by a series of refined experiments such as scanning tunneling microscopy (STM) down to 20 mK \cite{Enayat2016}, angle-dependent specific heat \cite{Kittaka2016}, London penetration depth \cite{Pang2016,Takenaka2017,Yamashita2017}, thermal conductivity \cite{Yamashita2017}, and angle-resolved photoemission spectroscopy (ARPES) \cite{Wu2021}. A multiband picture was soon developed based on first-principles electronic band-structure calculations, claiming the importance of both electron and hole Fermi surfaces \cite{Ikeda2015,Yang2018}. Different pairing candidates have since been proposed including the sign reversal $s^\pm$ wave due to a strong interband pairing interaction \cite{Yang2018}, the sign preserving $s^{++}$ wave \cite{Yamashita2017}, and the “$d+d$” mixed pairing motivated by the study of iron-pnictide superconductors \cite{Pang2016,Nica2017}. While all these proposals seem to fit well existing data of the specific heat  \cite{Kittaka2014,Kittaka2016,Yamashita2017} and the London penetration depth \cite{Pang2016,Yamashita2017}, a direct experimental probe of the gap structures is lacking due to the limited energy resolution. As a result, the exact gap structures of superconducting CeCu$_2$Si$_2$ at ambient pressure still remain undecided.

It is therefore crucial to look for smoking gun evidence that may distinguish these different scenarios. Among all possible probes, the impurity effect provides significant phase information on the pairing symmetry for both conventional and unconventional superconductors \cite{Zhu2006}. Actually, such experiments have been carried out on CeCu$_2$Si$_2$ shortly after its discovery and found that replacing Cu  by merely about 1\% Rh, Pd, or Mn \cite{Spille1983} can completely destroy the superconductivity. On the other hand, for  substitution of the Ce sites, the suppression of $T_c$ relies heavily on the dopant, with the critical concentration being 0.5\% for Sc, 6\% for Y, 10\% for La, and 20\% for Th \cite{Spille1983,Ahlheim1990}. These earlier measurements on the pair breaking effects caused by impurities were believed to provide strong support for nodal $d$-wave pairing in CeCu$_2$Si$_2$. But recent electron irradiation experiments reported surprising robustness of the superconductivity, thus hinting at a fully gapped $s^{++}$-wave pairing in contrast to similar measurements on cuprate and iron-pnictide superconductors with sign changing $d$- or $s^\pm$-wave gaps \cite{Yamashita2017}. On the other hand, the $s^\pm$-wave pairing has been used to explain the neutron spin resonance mode \cite{Stockert2011,Akbari2021}, the absence of the Hebel-Slichter peak, and the $1/T_1\propto T^3$ scaling of the spin-lattice relaxation rate \cite{Kitaoka1986,Fujiwara2008}. 

In this paper, we investigate the impurity effect in superconductor CeCu$_2$Si$_2$ and show that the above controversy may be resolved based on the fully gapped sign reversal $s^\pm$-wave pairing state \cite{Yang2018}. By using the Eliashberg gap functions and the $T$-matrix approach for a realistic two-band hybridization model, we present detailed theoretical studies on the suppression of $T_c$ and the density of states (DOS) with the intraband impurity scattering strength $u$ and the interband scattering strength $v$. We find a critical intraband scattering strength $U_{c}$ controlled by the ratio $r=v/u$ and the DOS at the Fermi energy, below which $T_c$ is quickly suppressed with increasing impurity concentration, while beyond which the superconductivity gradually revives with increasing $u$ and remains robust against impurities in the unitary limit. Such unexpected anomalous features reconcile the seeming discrepancy of the impurity effect induced by atomic substitution and electron irradiation, and hence provide important support for the fully gapped $s^\pm$ pairing in the heavy-fermion superconductor CeCu$_2$Si$_2$ at ambient pressure.

\begin{figure}[t]
\includegraphics[width=0.48\textwidth]{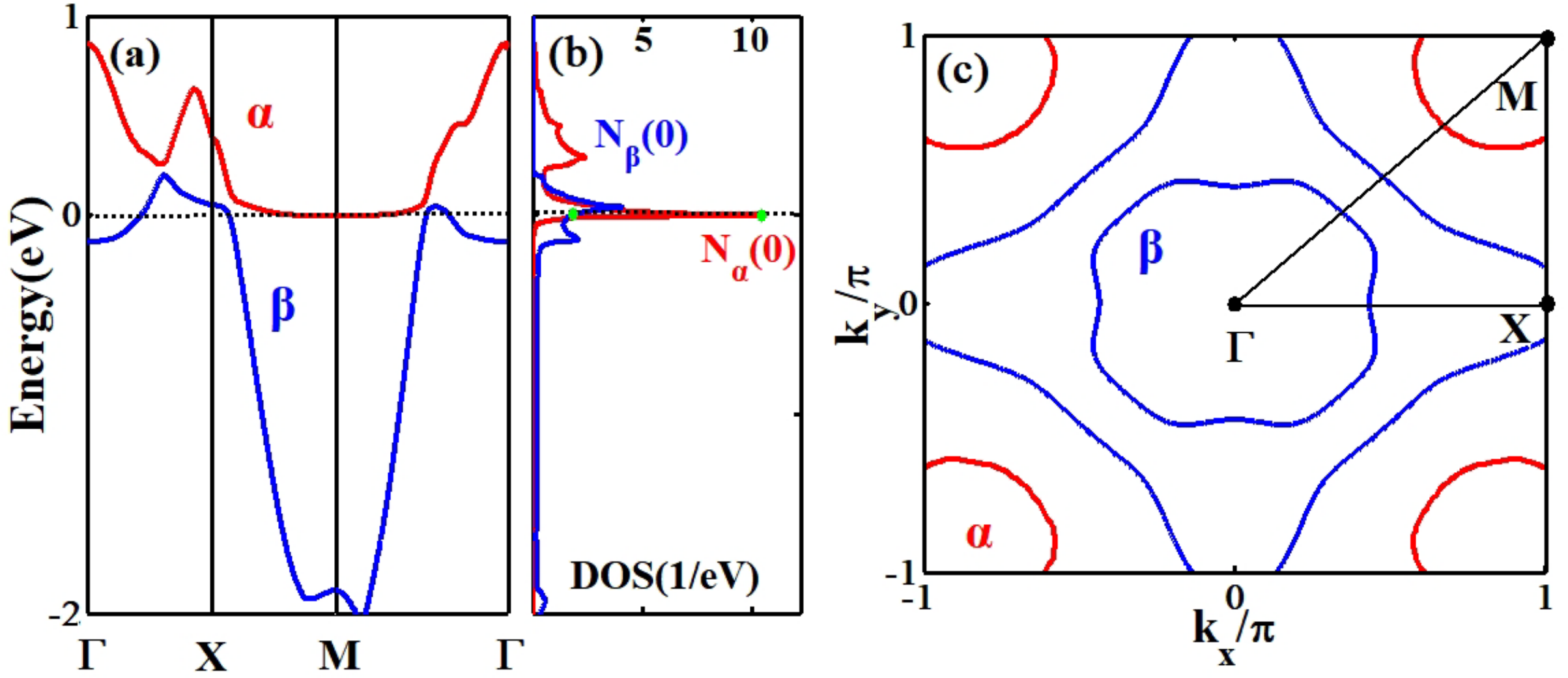}
\caption{(a) Electronic band structures of our effective two-band hybridization model from DFT+$U$ for CeCu$_2$Si$_2$ at ambient pressure. (b) Density of states of the $\alpha$ and $\beta$ bands in the normal state. $N_\alpha(0)$ and $N_\beta(0)$ refer to the DOS of $\alpha$ and $\beta$ bands at the Fermi energy. (c) A typical mapping of the two-dimensional Fermi surfaces with the hole Fermi surface $\beta$ and the heavy-electron Fermi surface $\alpha$.}
\label{fig1}
\end{figure}

Our effective two-band hybridization model \cite{Liu2019,Liu2022} is constructed based on density functional theory calculations (DFT+$U$), where only two hybridization bands cross the Fermi energy and dominate the low-energy physics of CeCu$_2$Si$_2$ \cite{Yang2018}. The electronic band structures are plotted in Fig. \ref{fig1}(a). A sharp peak is clearly seen in the DOS in Fig. \ref{fig1}(b), reflecting the extremely flat heavy electron band ($\alpha$) of typical $f$ character due to the many-body Kondo effect. At the Fermi energy, the DOS of the $\alpha$ band $N_\alpha(0)$ is about six times larger than that of the $\beta$ band $N_\beta(0)$. For a strong interband pairing interaction \cite{Yang2018}, this predicts a gap ratio $\mid\Delta^{\beta}/\Delta^{\alpha}\mid\approx [N_\alpha(0)/N_\beta(0)]^{1/2}\approx 2.4$, in good agreement with experimental estimates \cite{Kittaka2014,Enayat2016,Pang2016,Kittaka2016}. A typical two-dimensional mapping in Fig. \ref{fig1}(c) shows multiple Fermi surfaces, which fit well the recent ARPES measurement \cite{Wu2021}.

Without impurities, the linearized Eliashberg equation \cite{Dolgov2009,Allen1982} is written as 
\begin{eqnarray}
\Delta^{i}(i\omega_{n})=-\pi T \sum_{j=\alpha,\beta}g_{ij}N_{j} \sum_{m}{}^{'}\frac{\Delta^{j}(i\omega_{m})}{\left | \omega_{m} \right | },
\end{eqnarray}
where $i$, $j$ denote the band indices, $N_{j}$ refers to the DOS of band $j$ at the Fermi energy, $\omega_{n/m}$ is the fermionic Matsubara frequency, and $\sum_{m}^{'}\equiv \sum_{m}\theta(\omega_{c}-|\omega _{m}|)$, where $\omega_c$ is the cutoff. Clearly, the $s^\pm$-wave pairing is favored for a sufficiently large interband pairing interaction $g_{\alpha\beta}=g>0$. To focus on the impurity effect, we have neglected for simplicity the mass enhancement and quasiparticle damping induced by an electron-electron interaction. 

The effect of impurities can be well captured by the $T$-matrix approach as long as the concentration $n_{imp}\ll1$. The normal and anomalous self-energies are then given respectively by
\begin{eqnarray}
\textstyle \sum_{i}^{A}(i\omega_{n})&=&n_{imp}\sum_{j=\alpha,\beta }T_{ij}(i\omega_{n})f_{j}(i\omega_{n})T_{ij}(-i\omega_{n}),
\nonumber\\
\textstyle \sum_{i}^{N}(i\omega_{n})&=&n_{imp}T_{ii}(i\omega_{n}),
\label{eq2}
\end{eqnarray}
where $f_{j}(i\omega_{n})=\pi N_{j}\Delta^{j}(i\omega_{n})/|\omega_{n}|$ represents the local anomalous Green's function, $T_{ij}(i\omega_{n})=U_{ij}+\sum_{l=\alpha,\beta}U_{il}g_{l}(i\omega_{n})T_{lj}(i\omega_{n})$ is the $T$-matrix element for the nonmagnetic impurity scattering $U_{ij}=u\delta_{ij}+v(1-\delta_{ij})$, and $g_{l}(i\omega_{n})=-i\pi N_{l}\text{sgn}(\omega_{n})$ is the local normal Green's function \cite{Allen1982}. Taking into account these impurity effect, the linearized gap equation becomes \cite{Allen1982}
\begin{equation}
Z_{i}(i\omega_{n})\Delta^{i}(i\omega_{n})=-\pi T \sum_{j=\alpha,\beta}g_{ij}N_{j} \sum_{m}{}^{'}\frac{\Delta^{j}(i\omega_{m})}{|\omega_{m}|} + {\textstyle\sum_{i}^{A}}(i\omega_{n}),
\label{eq3}
\end{equation}
with the renormalization function
\begin{eqnarray}
Z_{i}(i\omega_{n})=1-\frac{{\textstyle \sum_{i}^{N}(i\omega_{n})}-{\textstyle \sum_{i}^{N}(-i\omega_{n})}}{2i\omega_{n}}.
\end{eqnarray}

When only the interband pairing interaction is considered for the case of dominant $s^\pm$-wave pairing, we obtain the following equation for $T_c$
\begin{eqnarray}
g^{2}N_{\alpha }N_{\beta }(X-B)(X-C)-\frac{2gN_{\alpha}N_{\beta}}{N_{\alpha }+N_{\beta }}(C-B)=1,
\end{eqnarray}
with
\begin{eqnarray}
X&=&\psi\left(\frac{\omega_{c}}{2\pi T_{c}}+1\right),\nonumber
B=\psi\left(\frac{1}{2}\right),\nonumber\\
C&=&\psi\left(\frac{1}{2}+(N_{\alpha }+N_{\beta})\frac{n_{imp}v^{2}}{2T_{c}D}\right),
\end{eqnarray}
where $D=1+\pi^{4}N_{\alpha}^{2}N_{\beta}^{2}(u^{2}-v^{2})^{2}+\pi^{2}u^{2}(N_{\alpha }^{2}+N_{\beta }^{2})+2\pi^{2}v^{2}N_{\alpha}N_{\beta}$ and $\psi(x)$ is the digamma function. For $n_{imp}\ll1$, the above equation may be approximately solved and gives
\begin{eqnarray}
\ln\frac{T_{c0}}{T_{c}}&\approx&\gamma \left[\psi\left(\frac{1}{2}+\frac{(N_{\alpha }+N_{\beta})n_{imp}v^{2}}{2T_{c}D}\right)-\psi\left(\frac{1}{2}\right)\right],\nonumber\\
\label{EqTc}
\end{eqnarray}
where $T_{c0}\sim0.6$ K is the superconducting transition temperature without impurities and $\gamma=1/2+\sqrt{N_\alpha N_\beta}/(N_\alpha+N_\beta)$. This expression is similar to previous results \cite{Eremin20081,Senga2009,Allen1982} where $T_c$ may be solved iteratively.

\begin{figure}[t]
\centering\includegraphics[width=0.48\textwidth]{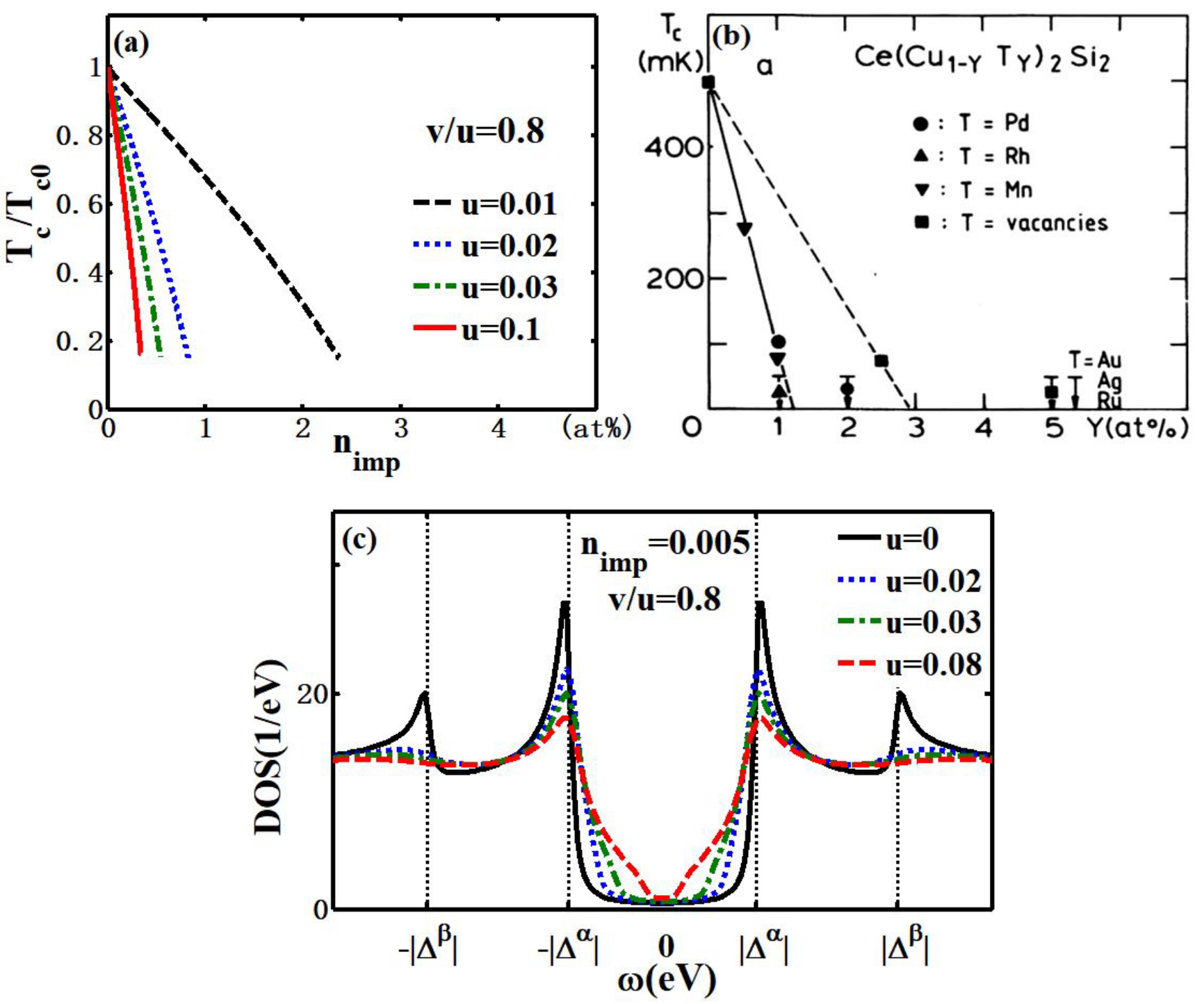}
\caption{(a) The calculated $T_{c}$/$T_{c0}$ as a function of the impurity concentration $n_{imp}$ for weak intraband impurity scattering  $u$ at a fixed ratio $v/u=0.8$. (b) Variation of $T_c$ of CeCu$_2$Si$_2$ with Cu substitutions in experiment, adapted from Ref.[19]. (c) Impurity-induced DOS for different values of $u$ at $n_{imp}=0.005$ and $v/u=0.8$.}
\label{fig2}
\end{figure}

For comparison with experiments, we plot in Fig. \ref{fig2}(a) the variation of $T_{c}$/$T_{c0}$ as a function of the impurity concentration $n_{imp}$ for a typical ratio $v/u=0.8$ in the Born or weak impurity scattering region (small $u$). As expected, $T_c$ drops quickly with $n_{imp}$ and the overall suppression is enhanced as the scattering potential $u$ increases. This provides a potential explanation of earlier doping experiments reproduced in Fig. \ref{fig2}(b) \cite{Spille1983}, where the superconductivity is destroyed in  similar manners for 1\% of Rh, Pd, or Mn replacement of Cu. Since Cu 3$d$ orbital contributes little component of the Fermi surface evidenced by DFT+$U$ calculations\cite{Wu2021}, its substitution is reasonably assumed to cause weak impurity scattering. Besides the suppression of $T_{c}$, the impurities also induce in-gap states. Figure \ref{fig2}(c) plots the DOS in the superconducting state derived from the imaginary part of the local Green's function for $n_{imp}=0.005$. In contrast to two intragap resonance peaks for a single nonmagnetic impurity \cite{Liu2019,Liu2022,Matsumoto2009}, an impurity band appears so that the original nearly U shaped DOS ($u=0$) turns into a V shape at $u=0.03$ and is almost filled in beyond $u=0.08$. This DOS variation might explain the power-law behavior in $1/T_{1}$ in earlier NMR measurement and the larger full gap and smaller nodal gap in tunneling spectra \cite{Enayat2016}.

\begin{figure}[t]
\centering\includegraphics[width=0.48\textwidth]{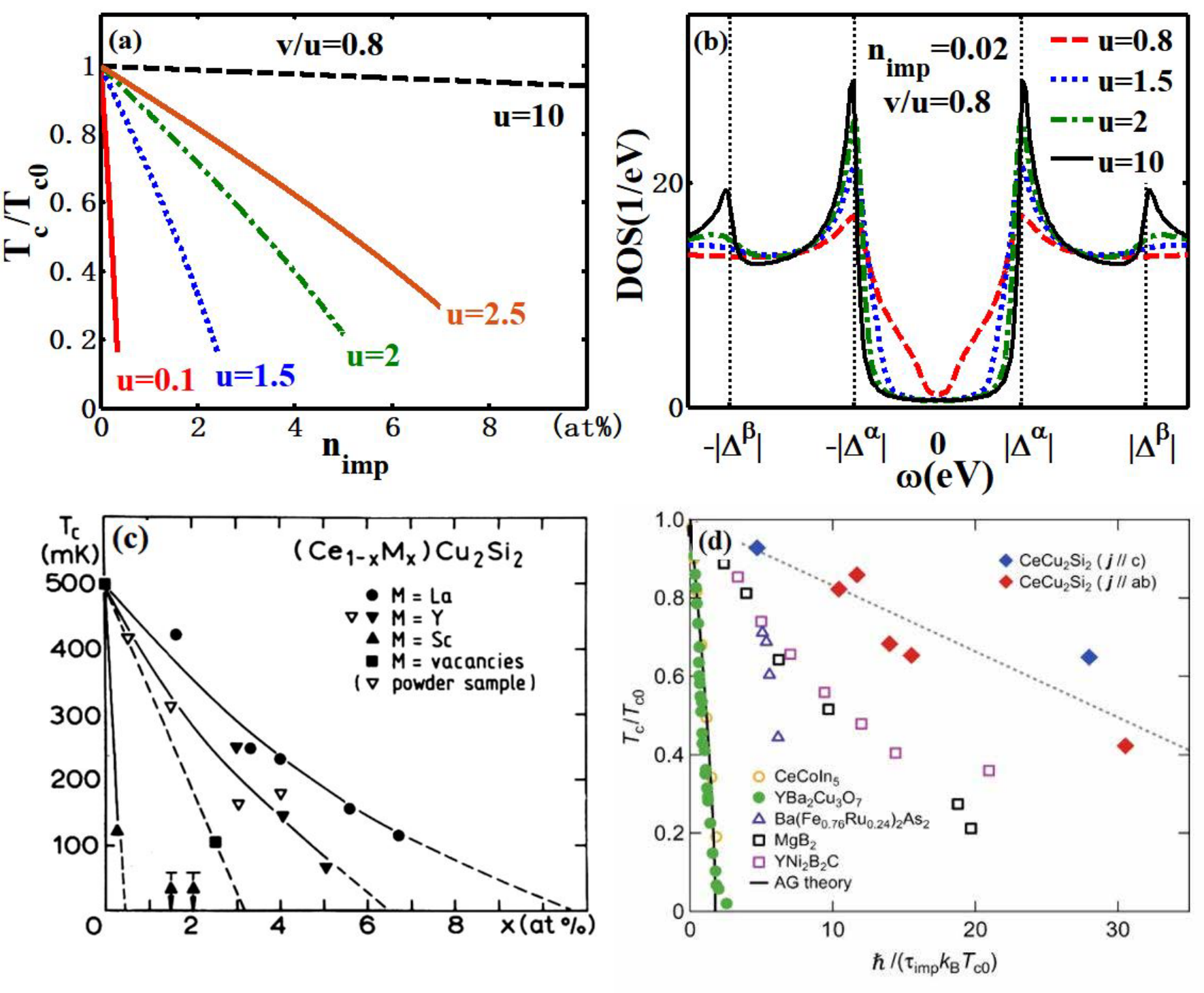}
\caption{(a) The calculated $T_{c}$/$T_{c0}$ as a function of the impurity concentrations $n_{imp}$ for strong intraband impurity scattering  $u$ at a fixed ratio $v/u=0.8$. (b) Impurity-induced DOS for different values of $u$ at $n_{imp}=0.02$ and $v/u=0.8$. (c) Variation of $T_c$ of CeCu$_2$Si$_2$ with Ce substitutions adapted from Ref.[19]. (d) Comparison of electron irradiation measurements on $T_c/T_{c0}$ in CeCu$_2$Si$_2$ and other superconductors adapted from Ref.[13].}
\label{fig3}
\end{figure}

So far it looks normal, but surprises come from the substitution or removal of Ce sites. Since Ce 4$f$ electrons dominate the superconductivity in CeCu$_2$Si$_2$, one may naively assume stronger impurity effect than those of Cu replacement. We have therefore perform further calculations for strong impurity scattering (large $u$). The resulting $T_{c}$/$T_{c0}$ is plotted in Fig. \ref{fig3}(a) as a function of $n_{imp}$ for the same ratio $v/u=0.8$. Quite remarkably, instead of further suppressing $T_c$ as in Fig. \ref{fig2}(a), the superconductivity seems to gradually revive with increasing $u$ and is robust against the impurities in the unitary limit ($u=10$). This suggests that very strong scattering potential (large $u$) has actually a weaker effect on $T_c$. Indeed, the critical concentrations for replacing Ce are 0.5\% Sc, 6\% Y, 10\% La, or 20\% Th in experiment as reproduced in Fig. \ref{fig3}(c) \cite{Spille1983,Ahlheim1990}. In order to make a direct comparison with experimental data, in Fig. \ref{fig3}(a) we estimate the scattering potential $u=0.1$ for Sc, $u=2$ for Y, and $u=2.5$ for La, respectively, since the scattering strength is believed to be in direct proportion to the ionic radius and the scattering cross section. In particular, removing some Ce ions by electron irradiation is expected to bring much stronger scattering than atomic substitution \cite{Yamashita2017}. But as shown in Fig. \ref{fig3}(d) adapted from Ref.[13], its suppression on $T_{c}$ is much slower than those in cuprate and iron-pnictide superconductors, but similar to conventional $s$-wave superconductors. Thus, the consistency between our theory and the experiments provides a potential explanation to the puzzling observation in the doping and electron irradiation experiments in CeCu$_2$Si$_2$, and consequently supports the nodeless $s^\pm$-wave pairing of its superconductivity. With increasing $u$, as shown in Fig. \ref{fig3}(b) for $n_{imp}=0.02$, the DOS recovers from a V shape at $u=0.8$ to a U shape at $u=10$, reflecting the robustness of superconductivity even in the unitary limit. It should be noted that the exact strength of the scattering potential for each type of impurity is not known in CeCu$_2$Si$_2$. It will be important if future materials calculations or refined experiments could clarify this aspect in order to fully establish the above scenario.

\begin{figure}[tb]
\centering\includegraphics[width=0.48\textwidth]{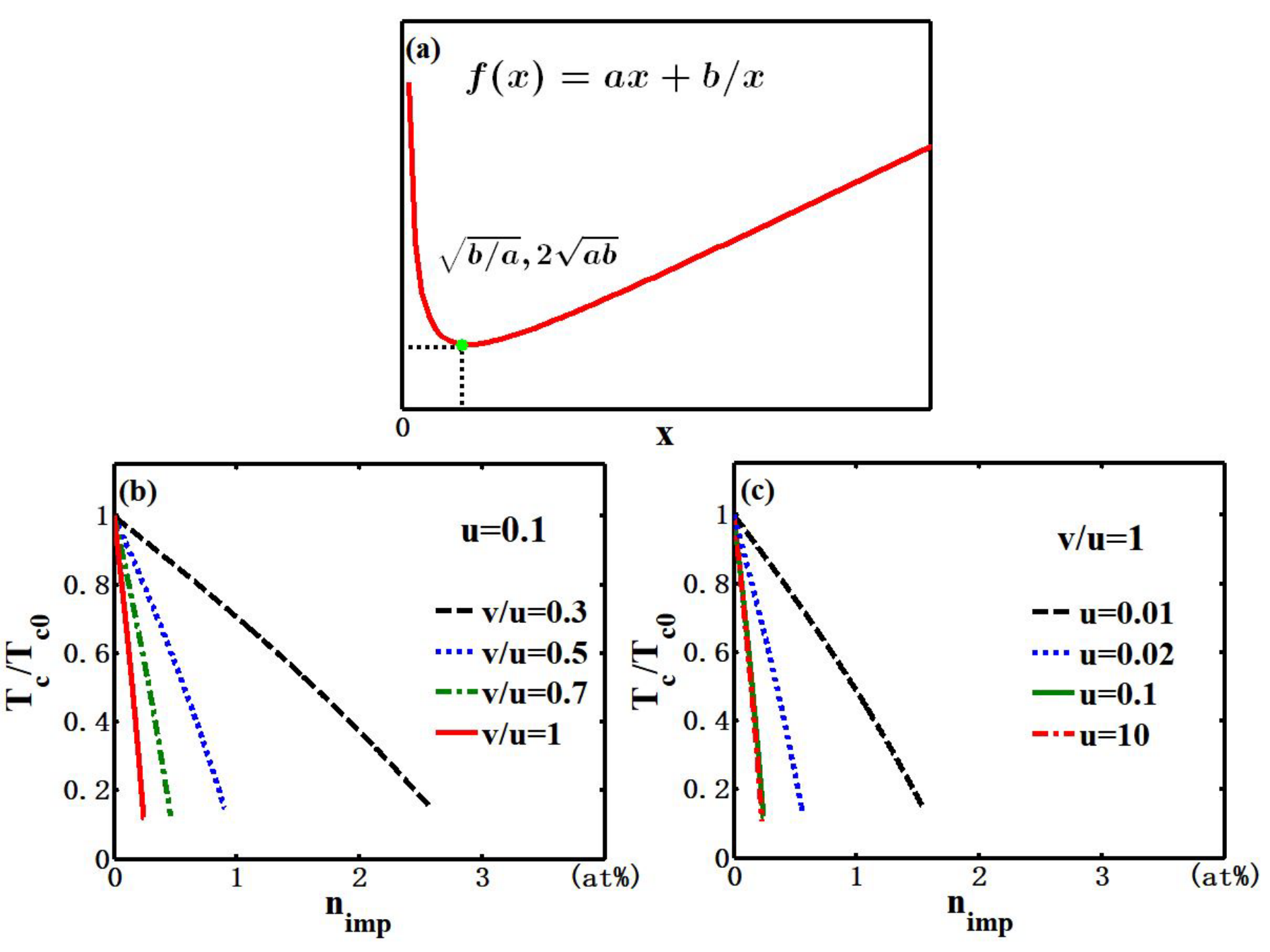}
\caption{(a) The Hook function of the standard form, $f(x)=ax+b/x$, with a crossing point at ($\sqrt{b/a},2\sqrt{ab}$). (b) $T_{c}$/$T_{c0}$ as a function of the impurity concentration $n_{imp}$ with varying ratio $v/u$ at $u=0.1$. (c) $T_{c}$/$T_{c0}$ as a function of $n_{imp}$ with varying intraband impurity scattering strength $u$ at $v/u=1$.}
\label{fig4}
\end{figure}

The anomalous behavior of $T_c$ suppression may be understood analytically from Eq. (\ref{EqTc}), where the difference of two digamma functions is fully determined by the magnitude of $(N_{\alpha }+N_{\beta})n_{imp}v^{2}/2T_{c}D$. This quantity increases linearly with $n_{imp}$, causing the monotonic suppression of $T_c$ with increasing impurity concentration. On the other hand, its evolution with the impurity scattering potential is more complicated because $D$ also depends on $u$ and $v$. To see its effect more clearly, we first fix the ratio $r\equiv v/u$ and obtain $(v^2/D)^{-1}=\pi^{4}N_{\alpha}^{2}N_{\beta}^{2}(1/r-r)^{2}u^{2}+1/r^{2}u^{2}+\pi^{2}(N_{\alpha}^{2}+N_{\beta}^{2})/r^{2}+2\pi^{2}N_{\alpha}N_{\beta}$, which is a typical Hook function of $u^2$ of the standard form $f(x)=ax+b/x$ up to a constant offset. As shown in Fig. \ref{fig4}(a), the Hook function contains a crossing point ($\sqrt{b/a},2\sqrt{ab}$). We have thus a critical impurity scattering strength $U_{c}^{2}=1/\pi^{2}N_{\alpha}N_{\beta}(1-r^{2})$ for $r<1$ ($v<u$), which is controlled by the DOS at the Fermi energy of both bands as well as the ratio $r$ between interband and intraband impurity scattering potentials. As a result, $T_c$ behaves opposite across $U_c$. For $u<U_{c}$, it is more strongly suppressed as $u$ increases, but for $u>U_{c}$, the superconductivity gradually revives and is robust against impurities in the unitary region. Physically, $v^2/D$ may be regarded as the renormalized effective interband scattering, whose  nonmonotonic dependence with the bare $v$ reflects the screening of interband scattering for sufficiently large $v$.

To have a complete understanding of the interband impurity scattering effect, we plot in Fig. \ref{fig4}(b) the $T_{c}$/$T_{c0}$ as a function of $n_{imp}$ with varying $r=v/u$ at fixed $u=0.1$. It is seen that $T_{c}$/$T_{c0}$ drops very quickly with increasing $r$, indicating that the interband scattering plays a key role in destroying the superconductivity for the nodeless $s^\pm$-wave pairing. However, no anomalous behavior is seen with increasing $r$ up to $r=1$. This can also be understood from the Hook function by rewriting $(v^2/D)^{-1}=\pi^{4}N_{\alpha}^{2}N_{\beta}^{2}u^{2}r^2+[\pi^{4}N_{\alpha}^{2}N_{\beta}^{2}u^{2}+\pi^{2}(N_{\alpha}^{2}+N_{\beta}^{2})+u^{-2}]/r^{2}-2\pi^{4}N_{\alpha}^{2}N_{\beta}^{2}u^{2}+2\pi^{2}N_{\alpha}N_{\beta}$. Fixing the impurity scattering $u$, we obtain a critical $r_{c}^{2}=\sqrt{1+(N_{\alpha}^{2}+N_{\beta}^{2})/\pi^{2}u^{2}N_{\alpha}^{2}N_{\beta}^{2}+1/\pi^{4}u^{4}N_{\alpha}^{2}N_{\beta}^{2}}$, which is always larger than unity and explains the absence of superconductivity revival in Fig. \ref{fig4}(b).

We note that there are two special cases: $v=0$ and $v=u$. In the first case, we always have $v^2/D=0$ and no suppression of $T_c$ occurs for any value of $u$. This implies that pure intraband impurity scattering cannot suppress the nodeless $s^\pm$-wave pairing, a result in good accordance with Anderson's famous theorem \cite{Anderson1959}. The case $v=u$ has been used to study the impurity effect in iron-pnictide and other multiband superconductors \cite{Parker2008,Bang2009}, but not suitable for CeCu$_2$Si$_2$. Figure \ref{fig4}(c) shows the results for $v/u=1$. We see that $T_{c}$ is always suppressed and drops quickly for large impurity scattering. This can be naturally understood from $(v^2/D)^{-1}=1/v^2+\pi^2(N_\alpha+N_\beta)^2$ at $r=1$, which decreases monotonically with increasing $v$ and approaches a constant in the unitary limit.

Given the general feature of the Hook function, one may wonder if the above anomalous property should be quite generally observed in all superconductors with nodeless $s^\pm$-wave pairing. This is, however, not the case. In fact, as shown already in Fig. \ref{fig3}(d) for Ba(Fe$_{0.76}$Ru$_{0.24}$)$_{2}$As$_{2}$, iron-pnictide superconductors also exhibit similar $s^\pm$-wave pairing but show more rapid $T_c$ suppression in electron irradiation experiments than CeCu$_2$Si$_2$ (though much weaker than cuprates). To understand this, we note that the critical $U_c$ derived above is strongly influenced by the magnitude of $(N_\alpha N_\beta)^{-1/2}$. Since the density of states $N_i$ is inversely proportional to the quasiparticle effective bandwidth, $U_c$ is tentatively given by the geometric average of two effective bandwidths, which is typically the order of $\sim0.1$ eV in heavy-fermion systems but much higher in other multiband superconductors. For iron pnictides, first-principles calculations yield a total DOS per Fe of about 1.3 eV$^{-1}$ per spin. The critical impurity scattering strength $U_{c}$ may then be estimated to be of the order of 1 eV at $r=0.8$, which is similar to the impurity potential 1.52 eV from Co substitution \cite{Kemper2009,Singh2008}. Thus, while its $T_c$ suppression may be weaker than $d$-wave superconductors, it is not yet fully in a region for the observation of the anomalous revival of $T_c$ as in CeCu$_2$Si$_2$. Our explanation is different from an earlier proposal of a possibly disorder-induced pairing transition from $s^\pm$ to $s^{++}$-wave pairing \cite{Efremov2011}. On the other hand, we speculate that this effect should be quite generally present in narrow-band systems with nodeless $s^\pm$-wave pairing, which might be a potential way to distinguish different pairing symmetries in, e.g., twisted bilayer graphene \cite{Yang2019} or the recently discovered bilayer or trilayer Ni-based superconductors \cite{Yang2023,Wang2023,Wang2024}.

To summarize, by analyzing the Eliashberg equations for a realistic two-band hybridization model and applying the $T$-matrix approach for its impurity effect, we reveal a critical scattering strength $U_{c}$ in the presence of interband impurity scattering and nodeless $s^\pm$-wave pairing. When the intraband impurity scattering $u<U_{c}$, superconductivity is quickly suppressed as a function of the impurity concentration, while above $U_{c}$, superconductivity revives and is found to be robust against impurities. This reconciles the controversy between doping and electron irradiation experiments, and provides important support for the fully gapped $s^\pm$-wave pairing in superconducting CeCu$_2$Si$_2$ at ambient pressure. 

We acknowledge fruitful discussions with Prof. Huaiming Guo, Prof. Li Ma, and Prof. Yu Lan. This work was supported by the National Natural Science Foundation of China (Grants No. 11774025, No. 11774401, and No. 12274036), the Strategic Priority Research Program of the Chinese Academy of Sciences (Grant No. XDB33010100), and the National Key Research and Development Program of China (Grant No. 2023YFA1406500, No. 2022YFA1402203, and No. 2021YFA1401803).

\end{document}